\documentclass[epj]{svjour}
\usepackage{graphicx,wrapfig,lipsum}
\usepackage{amssymb,amsmath}
\usepackage{subeqnarray}
\usepackage[usenames,dvipsnames]{color}
\begin{document}
\title{Lepton Number and Expansion of the Universe}
\author{Cheng Tao Yang${}^a$, Jeremiah Birrell${}^b$, Johann Rafelski${}^a$}
\institute{${}^a$Department of Physics, The University of Arizona, Tucson, Arizona 85721, USA\\
${}^b$Department of Mathematics and Statistics, University of Massachusetts Amherst, Amherst, MA 01003, USA}
\date{Version 2 of January 15, 2019; printed \today}

%%%%%%%%%%%%%%%%%%%%%%%%%%%%%%%%%%%%%%%%%%%%%%%%%%%%%%%%%%%%%%%%%%

\abstract{We study in a quantitative manner the lepton asymmetry $L$ (net leptons number per photon) in the Universe and its impact on Universe expansion. This is quantified via the standard tool, the effective number of Dirac neutrinos $N_\nu^{\mathrm{eff}}$. Specifically, we present the dependence of $N_\nu^{\mathrm{eff}}$ on the ambient temperature and the neutrino chemical potential and show the resultant neutrino yields in the cosmic background. We then obtain the dependence of $|L|$ and the baryon-number to lepton-number ratio, $B/|L|$, on $N_\nu^{\mathrm{eff}}$. Our results suggest that a value $|L|=0.46\pm0.06 $ and $B\simeq (1.33\pm 0.18)\times10^{-9}|L|$ may be considered. } 
%%%%%%%%%%%%%%%%%%%%%%%%%%%%%%%%%%%%%%%%%%%%%%%%%%%%%%%%%%%%%%%%%%%
\PACS{{98.80.Es} {Observational cosmology} {11.30.Fs} {Global symmetries} {98.70.Vc} {Background radiations} }
\maketitle
%%%%%%%%%%%%%%%%%%%%%%%%%%%%%%%%%%%%%%%%%%%%%%%%%%%%%%%%%%%%%%%%%%%
%%%%%%%%%%%%%%%%%%%%%%%%%%%%%%%%%%%%%%%%%%%%%%%%%%%%%%%%%%%%%%%%%%%

\section{Introduction}\label{Intro}
In the current standard cosmological model, asymmetry between leptons and anti-leptons $L\equiv [N_\mathrm{L}-N_{\overline{\mathrm{L}}}] /N_\gamma $ (normalized with the photon number) is in general assumed to be small (nano-scale) so that the net normalized lepton number equals the net baryon number $L=B$, $B=[N_\mathrm{B}-N_{\overline{\mathrm{B}}}]/N_\gamma $. However, there is another \lq natural\rq\ choice $L\simeq 1$, making the net lepton number and net photon number in the Universe similar. 

The Dirac neutrino cosmic background can shelter such a large lepton asymmetry $ N_\mathrm{L}-N_{\overline{\mathrm{L}}}\propto N_\gamma $~\cite{Caramete:2013bua,Lesgourgues:2018ncw,Castorina:2012md,Mangano:2011ip,Kang:1991xa,Oldengott:2017tzj,Serpico:2005bc,Barenboim:2017dfq,Barenboim:2016shh}. Models of efficient leptogenesis \cite{Serpico:2005bc,Barenboim:2016shh,Liu:1993am,Lunardini:2000fy,Davidson:2008bu} have been explored and they show that a large lepton asymmetry is a possible cosmic situation. This motivates a discussion of the observational manifestations of such asymmetry. 

Additional motivation arises from the widely discussed tension existent in astronomical analysis of the current era ($z\simeq 1-4$) Hubble expansion rate ($H_0$)~\cite{Riess:2018,Riess:2018byc}, as compared to the value $H_\mathrm{CMB}$ obtained when extrapolated from the era of photon decoupling ($z\simeq 1000$) obtained from CMB Planck data analysis~\cite{Ade:2013zuv,Aghanim:2018eyx}. 

Barenboim, Kinney and Park~\cite{Barenboim:2017dfq,Barenboim:2016shh} note that the lepton asymmetry of the Universe is one of the most weakly constrained parameters and propose leptogenesis scenarios able to accommodate a large lepton number asymmetry surviving up to this date. Our work extends their qualitative discussion of these constraints by quantifying the impact of large lepton asymmetry on Universe expansion.

Like prior work\cite{Kang:1991xa,Oldengott:2017tzj,Serpico:2005bc,Barenboim:2017dfq,Barenboim:2016shh,Liu:1993am,Lunardini:2000fy}, we assume that lepton asymmetry migrates into the Dirac neutrino background, creating different densities of neutrinos and antineutrinos. As is standard, the effects of such neutrino asymmetry on the expansion of the Universe is quantified via the effective number of neutrinos 
\begin{align}
\label{Neff}
N_\nu^{\mathrm{eff}}\equiv\frac{\rho^{\mathrm{tot}}_\nu}{\frac{7\pi^2}{120}\left(\frac{4}{11}\right)^{4/3}T_\gamma^4}\;,
\end{align}
where $\rho_\nu^{\mathrm{tot}}$ is the total energy density in neutrinos and $T_\gamma$ is the photon temperature. $N_\nu^{\mathrm{eff}}$ is defined such that three neutrino flavors with zero participation of neutrinos in reheating during $e^+e^-$ annihilation results in $N_\nu^{\mathrm{eff}}=3$. The factor of $\left(4/11\right)^{1/3}$ relates the photon temperature to the (effective) temperature of the free-streaming neutrinos after $e^\pm$ annihilation, under the assumption of zero neutrino reheating; see section~\ref{Effective_Number} below and e.g.~\cite{Kolb:1990vq,Birrell:2012gg} for further discussion.

The currently accepted theoretical value is $N_\nu^{\mathrm{eff}}=3.046$, after including the slight effect of neutrino reheating \cite{Mangano:2005cc,Birrell:2014uka}. The favored value of $N_\nu^{\mathrm{eff}}$ can be found by fitting to CMB data. In 2013 the Planck collaboration found $N_\nu^{\mathrm{eff}}=3.36\pm0.34$ (CMB only) and $N_\nu^{\mathrm{eff}}= 3.62\pm0.25$ (CMB and $H_0$)~\cite{Ade:2013zuv}; moreover, the discrepancy between $H_\mathrm{CMB}$ and $H_0$ has increased~\cite{Riess:2018,Riess:2018byc,Aghanim:2018eyx}. This tension, and the possibility that leptogenesis in the early Universe resulted in neutrino asymmetry, motivates our study of the dependence of $N_\nu^{\mathrm{eff}}$ on $L$.

Here and below we refer to yields (upper case subscript letters $N_\mathrm{L}, N_\mathrm{B},$ etc) in the comoving, {\it i.e.\/}, expanding, Universe volume, scaled with the Universe expansion factor $a(t)^3$ (FLRW cosmological metric $g_{00}=1,\;g_{ii}=-a^2, i=1,2,3$), where $H\equiv \dot a/a$ is the Hubble expansion rate. The present epoch measured rate is denoted $H_0$ and the present rate extrapolated from $z\simeq 1000$ is denoted by $H_\mathrm{CMB}$. The scaling factor is also often found in the temperature parameter as we discuss below. Lower letters {\it i.e.\/}, $n_\nu$, refer to the time-dependent (here neutrino) particle number density in the (FLRW) Universe rest frame.

In section~\ref{Effective_Number} we introduce a chemical potential that allows for asymmetry between neutrinos and antineutrinos and derive the relation between the effective number of neutrinos $N_\nu^{\mathrm{eff}}$ and the neutrino chemical potential. In section~\ref{Chemical_Potential} we relate the neutrino chemical potential to the ratios $|L|$ and $B/|L|$, allowing us to study the dependence of $N_\nu^{\mathrm{eff}}$ on these, in principle, physical observables. The implications of these results will be discussed in section~\ref{Discussion}.
%%%%%%%%%%%%%%%%%%%%%%%%%%%%%%%%%%%%%%%%%%%%%%%%%%%%%%%%%%%%%%%%%%%
%%%%%%%%%%%%%%%%%%%%%%%%%%%%%%%%%%%%%%%%%%%%%%%%%%%%%%%%%%%%%%%%%%%

\section{Relation Between $N_\nu^{\mathrm{eff}}$ and Neutrino Chemical Potential}
\label{Effective_Number}
Neutrinos gradually decouple~\cite{Birrell:2014gea} at a temperature of $T_f\simeq 2\,\mathrm{MeV}$ and are subsequently free-streaming. Assuming exact thermal equilibrium at the time of decoupling, the neutrino distribution can be subsequently written as (see~\cite{Birrell:2012gg} and references therein)
\begin{align}
\label{fnudef}
&f_\nu=\frac{1}{\exp{\left(\sqrt{\frac{E^2-m_\nu^2}{T_\nu^2}+\frac{m^2_\nu}{T^2_f}}-\sigma\frac{\mu_\nu}{T_f}\right)+1}}\;,
\end{align}
where $\sigma=+1(-1)$ denotes particles (antiparticles) and we define the effective neutrino temperature $T_\nu$ by
\begin{align}
T_\nu\equiv\frac{a(t_f)}{a(t)}T_f\;.
\end{align}
This definition of effective temperature for a free-streaming particle is motivated by the red-shifting of momentum in the comoving volume element of the Universe.

Since $T_f\gg m_\nu$ and also $T_\nu\gg m_\nu$ in the domain of our analysis, we can consider the massless limit in Eq.\;(\ref{fnudef}). The comoving energy density for neutrinos (antineutrinos) can then be written as
\begin{align}
\rho_{\nu(\bar{\nu})}=\frac{g_\nu}{2\pi^2}\int_0^\infty\frac{p^3\,dp}{\exp{\left(p/T_\nu\mp\mu_\nu/T_f\right)}+1}.
\end{align}
This yields the total neutrino-antineutrino energy density
\begin{align}
\rho_\nu^{\mathrm{tot}}=\frac{g_\nu}{2\pi^2}\left(\frac{T_\nu}{T_f}\right)^{\!\!4}\!\!\displaystyle\int_0^\infty\!\!\left(
\displaystyle\frac{\xi^3\,d\xi}{e^{\frac{\xi-\mu_\nu}{T_f}}+1}+
\displaystyle\frac{\xi^3\,d\xi}{e^{\frac{\xi+\mu_\nu}{T_f}}+1}\right),
\end{align}
where we made the change of variables $\xi\equiv (T_f/T_\nu) p$. The integral can be evaluated by using the formula \cite{Elze:1980er}
\begin{align}
\label{Integral}
I_n&\equiv\sum_\sigma\int_0^\infty\frac{p^n\sigma^{n+1}dp}{\exp[\frac{1}{T}\left(p-\sigma\mu\right)]+1}\notag\\
=\sum^{n-1}_{k=0}&\binom{n}{k}\,\mu^k\,T^{(n-k+1)}\left[1+(-1)^{n+k+1}\right]\left(1-2^{k-n}\right)\,\notag\\&\times\Gamma(n-k+1)\zeta(n-k+1)+\frac{\mu^{n+1}}{n+1},
\end{align}
where $\Gamma$ and $\zeta$ are the gamma function and Riemann zeta function respectively. Hence the total neutrino energy density (again, under the approximations $T_f\gg m_\nu$, $T_\nu\gg m_\nu$) can be written as
\begin{align}
\label{Energy_Density}
\rho_\nu^{\mathrm{tot}}&=\frac{g_\nu}{2\pi^2}\left(\frac{T_\nu}{T_f}\right)^{\!\!4}\,I_3\notag\\
&=\frac{g_\nu\,T_\nu^4}{2\pi^2}\left[\frac{7\pi^4}{60}+\frac{\pi^2}{2}\left(\frac{\mu_\nu}{T_f}\right)^{\!\!2}+\frac{1}{4}\left(\frac{\mu_\nu}{T_f}\right)^{\!\!4}\right].
\end{align}
Substituting Eq.\;(\ref{Energy_Density}) into Eq.\;(\ref{Neff}), the effective number of neutrinos is given by 
\begin{align}
\label{Neff_002}
N_\nu^{\mathrm{eff}}\!\!
=\!3\!\left(\frac{11}{4}\right)^{\!\!\frac{4}{3}}\!\!\left(\frac{T_\nu}{T_\gamma}\right)^{\!\!4}\!
\left[1\!+\!\frac{30}{7\pi^2}\!\!\left(\frac{\mu_\nu}{T_f}\right)^{\!\!2} 
\!\!+\frac{15}{7\pi^4}\!\!\left(\frac{\mu_\nu}{T_f}\right)^{\!\!4}\right].
\end{align}
From Eq.\;(\ref{Neff_002}) we obtain, for the standard photon reheating ratio $T_\nu/T_\gamma=(4/11)^{1/3}$ \cite{Kolb:1990vq} and degeneracy $g_\nu=3$ (flavor), the relation between the effective number of neutrinos and the chemical potential at freezeout \cite{Lesgourgues:2018ncw,Castorina:2012md,Mangano:2011ip,Kang:1991xa,Oldengott:2017tzj,Serpico:2005bc}
\begin{align}
\label{Neff_Potential}
N_\nu^{\mathrm{eff}}=3\left[1+\frac{30}{7\pi^2}\left(\frac{\mu_\nu}{T_f}\right)^{\!\!2}+ \frac{15}{7\pi^4} \left(\frac{\mu_\nu}{T_f}\right)^{\!\!4}\right]\;.
\end{align}
As noted, Eq.\;(\ref{Neff_Potential}) is well known. However, we present here for the first time a quantitative exploration of its consequences in the context of neutrino cosmology.

\begin{figure}[t]
\begin{center}
\includegraphics[width=3.3in]{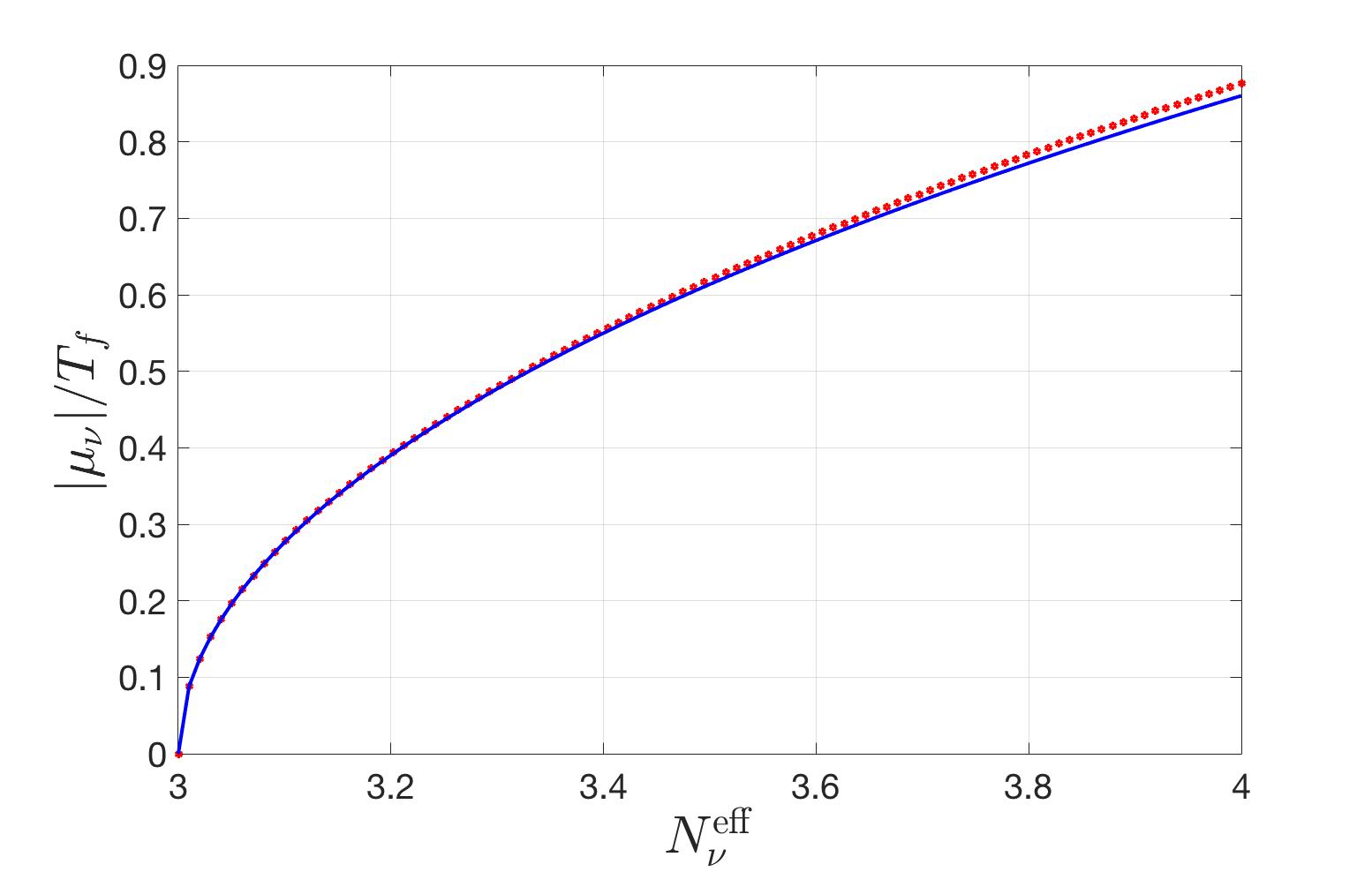}
\caption{The free-streaming neutrino chemical potential $|\mu_\nu|/T_f$ as a function of the effective number of neutrinos $N_\nu^{\mathrm{eff}}$. The solid (blue) line is the exact solution and the (red) dashed line is the approximate solution neglecting the $(\mu_\nu/T_f)^4$ term; the maximum difference in the domain shown is about $2\%$.}
\label{Chemical_Potential_Neff}
\end{center}
\end{figure}

In order to relate the chemical potential to $N_\nu^{\mathrm{eff}}$ we solve Eq.\;(\ref{Neff_Potential}) for $\mu_\nu/T_f$ and obtain 
\begin{align}
\frac{\mu_\nu}{T_f}=&\pm \pi\sqrt{\sqrt{1+\frac{7}{15}\left(\frac{N_\nu^{\mathrm{eff}}}{3}-1\right)}-1}\\
\approx&\pm\sqrt{\frac{7\pi^2}{30}\left(\frac{N_\nu^{\mathrm{eff}}}{3}-1\right)}.\label{Solution}
\end{align}
The approximation is the result of Taylor expanding the innermost root, or equivalently, from neglecting the $(\mu_\nu/T_f)^4$ term in Eq.\;(\ref{Neff_Potential}). In Fig.\;\ref{Chemical_Potential_Neff} we plot the free-streaming neutrino chemical potential $|\mu_\nu|/T_f$ as a function of the effective number of neutrinos $N_\nu^{\mathrm{eff}}$. In the parameter range of interest, we find that the term $(\mu_\nu/T_f)^4$ only contributes $\approx 2\%$ to the calculation and henceforth we neglect it, and use the approximation Eq.\;(\ref{Solution}). 

The SM value of the effective number of neutrinos, $N_\nu^{\mathrm{eff}}=3$, is obtained under the assumption that the neutrino chemical potentials are not essential, {\it i.e.\/}, $\mu_\nu\ll T_f$. From Fig.\;\ref{Chemical_Potential_Neff}, to interpret the literature values $N_\nu^{\mathrm{eff}}=3.36\pm0.34$ (CMB only) and $N_\nu^{\mathrm{eff}}= 3.62\pm0.25$ (CMB and $H_0$) we require $0.52\leqslant\mu_\nu/T_f\leqslant0.69$. These values suggest a possible neutrino-antineutrino asymmetry at freezeout, {\it i.e.\/} a difference between the number densities of neutrinos and antineutrinos.
%%%%%%%%%%%%%%%%%%%%%%%%%%%%%%%%%%%%%%%%%%%%%%%%%%%%%%%%%%%%%%%%%%
%%%%%%%%%%%%%%%%%%%%%%%%%%%%%%%%%%%%%%%%%%%%%%%%%%%%%%%%%%%%%%%%%%%

\begin{figure}[t]
\begin{center}
\includegraphics[width=3.3in]{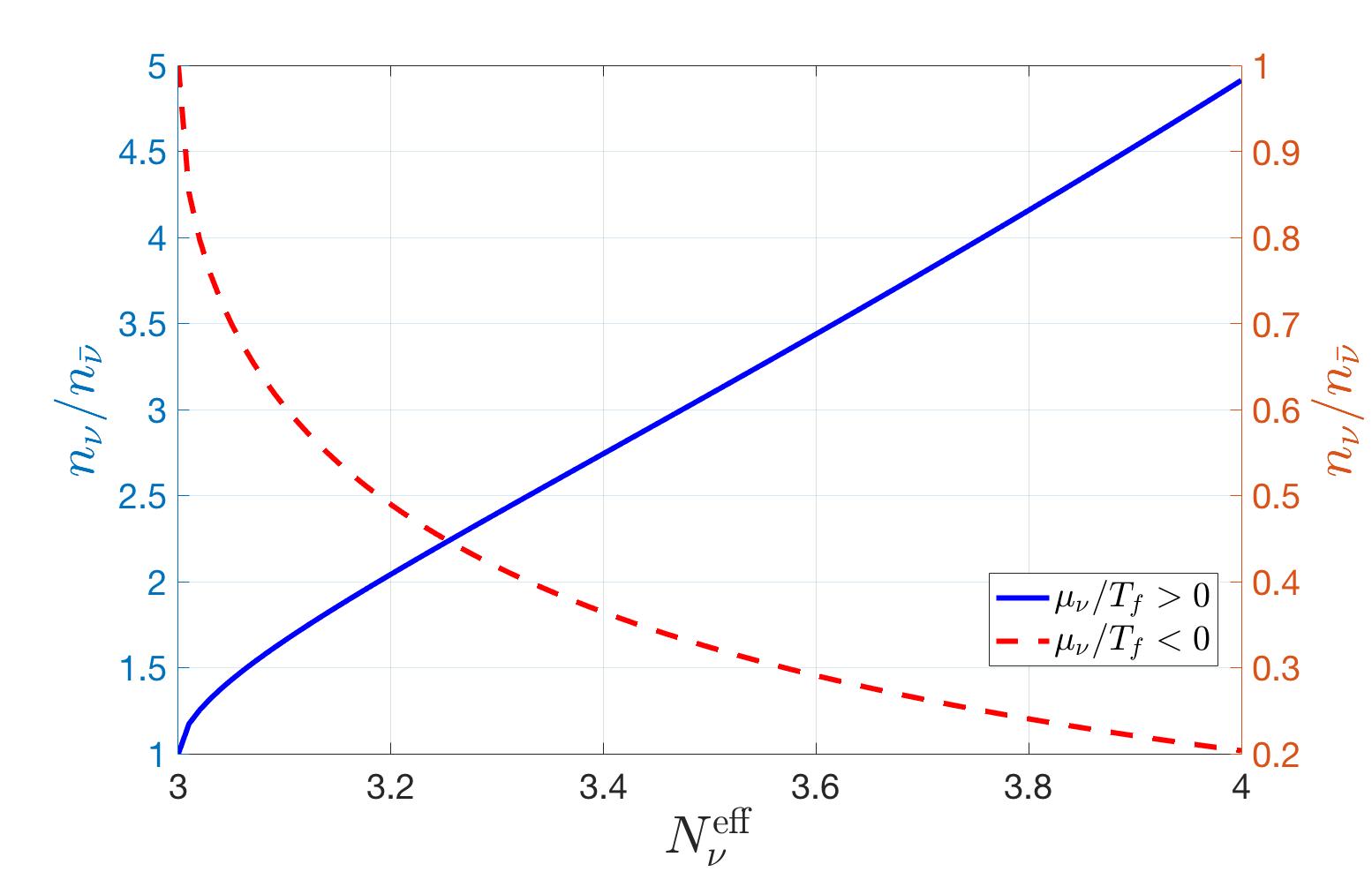}
\caption{The fixed (once the Universe cools below neutrino freezeout temperature $T_f\simeq 2\,\mathrm{MeV}$) ratio between neutrino and antineutrino number density as a function of the effective number of neutrino species $N_\nu^{\mathrm{eff}}$.}
\label{Neutrino_Density}
\end{center}
\end{figure}

The neutrino (antineutrino) number density is given by
\begin{align}\label{delta_n}
n_{\nu(\bar{\nu})}=\frac{g_\nu}{2\pi^2}\int\frac{p^2\,dp}{\exp{\left(p/T_\nu\mp\mu_\nu/T_f\right)}+1}\;,
\end{align}
where use lower letters for densities (no expansion factor $a^3$). The difference in yield of neutrinos and antineutrinos can evaluated using Eq.\;(\ref{Integral})
\begin{align}
\label{Excess_Neutrino}
n_\nu-n_{\overline{\nu}}=\frac{g_\nu}{6\pi^2}T^3_\nu\bigg[\pi^2\left(\frac{\mu_\nu}{T_f}\right)+\left(\frac{\mu_\nu}{T_f}\right)^{\!\!3}\bigg].
\end{align}
Hence the density ratio between neutrinos and antineutrinos can be written as
\begin{align}
\frac{n_\nu}{n_{\bar{\nu}}}&=\frac{1}{n_{\bar{\nu}}}\bigg[n_{\bar{\nu}}+\frac{g_\nu}{6}\left(\frac{\mu_\nu}{T_f}\right)T^3_\nu+\frac{g_\nu}{6\pi^2}\left(\frac{\mu_\nu}{T_f}\right)^{\!\!3}T^3_\nu\bigg]\notag\\
&=1+\frac{1}{3}\bigg[\pi^2 \tilde\mu_\nu+\tilde\mu_\nu^3\bigg]\left(\int^\infty_0\!\!\!\!\frac{\xi^2d\xi}{\exp{(\xi+\tilde\mu_\nu)}+1}\right)^{\!\!-1},
\end{align}
where we define the dimensionless variables $\xi=p/T_\nu$, $\tilde\mu_\nu=\mu_\nu/T_f$. In Fig.\;\ref{Neutrino_Density} we show the ratio between neutrino and antineutrino number density as a function of the effective number of neutrino species $N_\nu^{\mathrm{eff}}$. 

\begin{figure}[t]
\begin{center}
\includegraphics[width=3.3in]{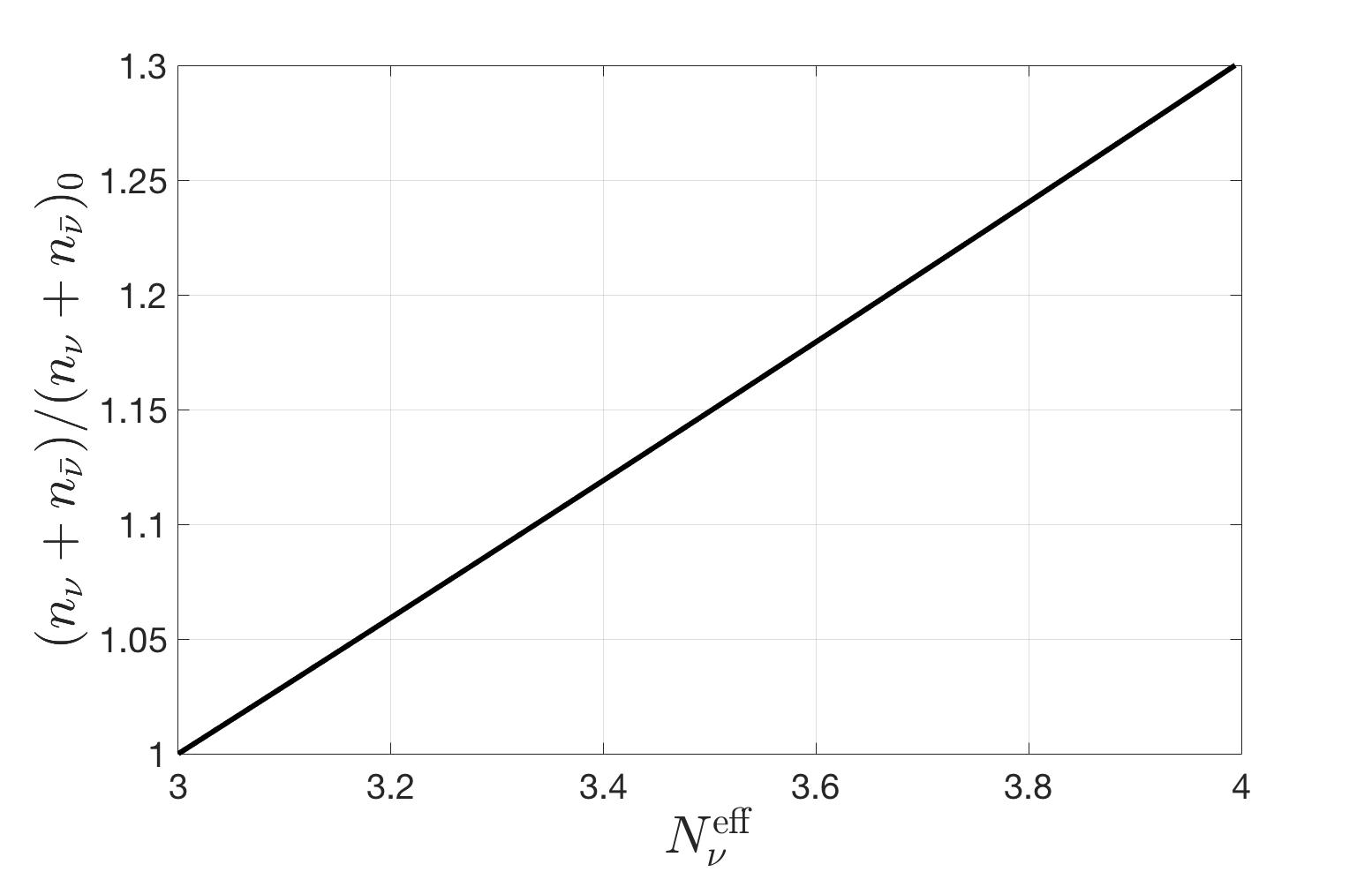}
\caption{The fixed (once the Universe cools below neutrino freezeout temperature $T_f\simeq 2\,\mathrm{MeV}$) ratio between the total neutrino number density $\big(n_\nu+n_{\bar{\nu}}\big)$ and the number density at zero chemical potential $\big(n_\nu+n_{\bar{\nu}}\big)_0$ as a function of the effective number of neutrino species $N_\nu^{\mathrm{eff}}$. }
\label{Density_tot}
\end{center}
\end{figure}

At zero chemical potential, we can also compute the reference total neutrino density 
\begin{align}
\big(n_\nu+n_{\bar{\nu}}\big)_0=2\times\left(\frac{3}{4}\frac{\zeta(3)}{\pi^2}g_\nu T^3_\nu\right).
\end{align}
Again, after changing variables $\xi=p/T_\nu$, the ratio between the total number density and number density at zero chemical potential can be written 
\begin{align}\label{n_ratio}
\frac{ n_\nu+n_{\bar{\nu}} }{\big(n_\nu+n_{\bar{\nu}}\big)_0} 
=\frac{1}{3\zeta(3)}&\left[
 \int\frac{\xi^2\,d\xi }{\exp{\left(\xi-\tilde\mu_\nu\right)}+1}
\right.\\
&+\left.\int\frac{\xi^2\,d\xi }{\exp{\left(\xi+\tilde\mu_\nu\right)}+1}\right].\notag
\end{align}
In Fig.\;\ref{Density_tot} we plot this ratio as a function of $N_\nu^{\mathrm{eff}}$. Note that invariance of Eq.\;(\ref{n_ratio}) under $\mu_\nu\to-\mu_\nu$ means that the cases $n_\nu>n_{\bar{\nu}}$ or $n_{\bar{\nu}}>n_\nu$ are indistinguishable here and both lead to $N_\nu^{\mathrm{eff}}>3$. Corresponding to the values $N_\nu^{\mathrm{eff}}=3.36\pm0.34$ (CMB) and $N_\nu^{\mathrm{eff}}= 3.62\pm0.25$ (CMB and $H_0$), we have $1.108\leqslant \big(n_\nu+n_{\bar{\nu}}\big)/\big(n_\nu+n_{\bar{\nu}}\big)_0 \leqslant1.187$, {\it i.e.\/}, the increase in the total number density is approximately $10.8\%$ to $18.7\%$ today.

%%%%%%%%%%%%%%%%%%%%%%%%%%%%%%%%%%%%%%%%%%%%%%%%%%%%%%%%%%%%%%%%%%%
%%%%%%%%%%%%%%%%%%%%%%%%%%%%%%%%%%%%%%%%%%%%%%%%%%%%%%%%%%%%%%%%%%%

%%%%%%%%%%%%%%%%%%%%%%%%%%%%%%%%%%%%%%%%%%%%%%%%%%%%%%%%%%%%%%%%%%%
%%%%%%%%%%%%%%%%%%%%%%%%%%%%%%%%%%%%%%%%%%%%%%%%%%%%%%%%%%%%%%%%%%%
%%%%%%%%%%%%%%%%%%%%%%%%%%%%%%%%%%%%%%%%%%%%%%%%%%%%%%%%%%%%%%%%%%%
\section{Dependence of $N_\nu^{\mathrm{eff}}$ on Lepton Asymmetry}
\label{Chemical_Potential}
We now obtain the relation between neutrino chemical potential and the baryon to lepton ratio. Let us consider the neutrino freezeout temperature $T_f\simeq 2.0$ MeV; here we treat neutrino freezeout as occurring instantaneously and prior to $e^\pm$ annihilation (implying zero neutrino reheating). Comoving lepton (and baryon) number is conserved after the epoch of leptogenesis (baryogenesis, respectively) which (as all agree) precedes the epoch under consideration in this work ($T\lesssim 2$\;MeV). However, the photon number $N_\gamma$ changes due to reheating, and accommodates the large number of photons originating from $e^+e^-$-annihilation. We present results in terms of the current epoch photon number. 

At energies of the order of a few MeV, we consider the Universe containing the following ingredients:
\begin{align}
&\textrm{Photon}: \gamma\notag\\
&\textrm{Baryon}: n, p\notag\\
&\textrm{Charged lepton}: e^-, e^+\notag\\
&\textrm{Neutrino}: \nu_e, \nu_\mu, \nu_\tau, \overline{\nu}_e,\overline{\nu}_\mu,\overline{\nu}_\tau\notag.
\end{align}
The neutrino chemical potentials at freezeout, which are imprinted into the free-streaming distribution Eq.\;(\ref{fnudef}), can be determined by the following criteria in the Universe \cite{Fromerth:2012fe}:
\begin{itemize}
\item Charge neutrality $(n_e-n_{\overline{e}})=(n_p)$; this condition is required to eliminate Coulomb energy and maintain charge neutrality in the Universe.
\item The total entropy-per-baryon is a constant, $s/n_B=$ Const, after neutrino freezeout. This follows from observing that both free-streaming systems and systems in equilibrium have preserved comoving entropy.
\end{itemize}

The lepton-density asymmetry $\ell $ at neutrino freezeout can be written as
\begin{align}
\ell_f \equiv\big(n_e-n_{\overline{e}}\big)_f+\sum_{i=e,\mu, \tau}\big(n_{\nu_i}-n_{\overline{\nu}_i}\big)_f,
\end{align}
where we use the subscript $f$ to indicate that the quantities should be evaluated at the neutrino freezeout temperature. As a first approximation, here we assume that all neutrinos freeze out at the same temperature and their chemical potentials are the same; {\it i.e.\/},
\begin{align}
\mu_\nu=\mu_{\nu_e}=\mu_{\nu_\mu}=\mu_{\nu_\tau}.
\end{align}
Furthermore neutrino oscillation implies that neutrino number is freely exchanged between flavors; {\it i.e.\/}, $\nu_e\rightleftharpoons\nu_\mu\rightleftharpoons\nu_\tau$, and we can assume that all neutrino flavors share the same population. Under these assumptions, the lepton-density asymmetry can be written as
\begin{align}
\label{L_asymmetry} 
\ell_f=\big(n_e-n_{\overline{e}}\big)_f+\big(n_{\nu}-n_{\overline{\nu}}\big)_f,
\end{align}
where the three flavors are accounted for by taking the degeneracy $g_\nu=3$ in the last term.

The baryon-density asymmetry $b$ at neutrino freezeout is given by
\begin{align}
\label{B_asymmetry}
b_f \equiv\big(n_p-n_{\overline{p}}\big)_f+\big(n_n-n_{\overline{n}}\big)_f \approx \big(n_p+n_n\big)_f,
\end{align}
where $n_{\overline{n}}$ and $n_{\overline{p}}$ are negligible in the temperature range we consider here. Taking the ratio $\ell_f/b_f$, using charge neutrality, and introducing the entropy density we obtain
\begin{align}\label{Lf_Bf}
\ell_f /b_f 
\approx\big(n_p/n_B \big)_f+\big(n_{\nu}-n_{\overline{\nu}}\big)_f \big( s/n_B\big)_f s_f^{-1},
\end{align}
where use lower letters for densities (no expansion factor $a^3$), for example $n_B$ for the baryon number density, and the proton concentration at neutrino freezeout is given by
\begin{align}
\label{X_proton}
(n_p/n_B)_f&=\frac{1}{1+(n_n/n_p)_f}\\
&=\frac{1}{1+\exp{\big[-\left(Q+\mu_\nu\right)/T_f\big]}},
\end{align}
with $Q=m_n-m_p=1.293\,\mathrm{MeV}$. We neglect the electron chemical potential in the last step because the $e^\pm$ asymmetry is determined by the proton density, and at energies of order a few MeV, the proton density is small, {\it i.e.\/}, $\mu_e\ll T_f$. 

However, as we will see, for our study of $N_\nu^{\mathrm{eff}}$ we will be interested in the case of a large lepton-to-baryon ratio. From Eq.\;(\ref{X_proton}) it is apparent that this can only be achieved through the second term in Eq.\;(\ref{Lf_Bf}), with the first term then being negligible, as it is less than $1$. So we further approximate
\begin{align}\label{L_B_ratio}
\ell_f /b_f 
 \approx \big(n_{\nu}-n_{\overline{\nu}}\big)_f \big( s/n_B\big)_f s_f^{-1}.
\end{align}
We retained the full expression Eq.\;(\ref{X_proton}) in our above discussion to show that the presence of a chemical potential $\mu_\nu\simeq 0.2 Q$ could lead to small, perhaps noticeable, effects on pre-BBN proton and neutron abundance. We defer this unrelated discussion to a separate work. 

Note that for large $|\mu_\nu|$, Eq.\;(\ref{L_B_ratio}) implies that the signs of $\mu_\nu$ and $\ell_f$ are the same. However, for very small $\mu_\nu$ the sign of $\ell_f$ is determined by the interplay between (anti)electrons and (anti)neutrinos; {\it i.e.\/}, there is competition between the two terms in Eq.\;(\ref{L_asymmetry}).

The total entropy density at freezeout can be written
\begin{align}
\label{Entropy_density}
s_f=\frac{2\pi^2}{45}g^s_\ast(T_f)\,T_f^3,
\end{align}
where the $g^s_\ast$ counts the degree of freedom for relativistic particles~\cite{Kolb:1990vq}. At $T_f\simeq 2\mathrm{MeV}$, the relativistic species in the early Universe are photons, electron/positrons, and $3$ neutrino species. We have
\begin{align}
g^s_{\ast}&= g_\gamma+\frac{7}{8}\,g_{e^\pm}+\frac{7}{8}\,g_{\nu\bar{\nu}}\left(\frac{T_\nu}{T_\gamma}\right)^{\!\!3}\bigg[1+\frac{15}{7\pi^2}\left(\frac{\mu_\nu}{T_f}\right)^{\!\!2}\bigg]\notag\\
&=10.75+\frac{45}{4\pi^2}\left(\frac{\mu_\nu}{T_f}\right)^{\!\!2}\;,
\end{align}
where the degrees of freedoms are given by $g_\gamma=2$, $g_{e^\pm}=4$, and $g_{\nu\bar{\nu}}=6$, and we have $T_\nu=T_\gamma=T_f$ at neutrino freezeout.
%%%%%%%%%%%%%%%%%%%%%%%%%%%%%%%%%%%%%%%%%%%%%%%%%%%%%%%%%%%%%%%%%%%
\begin{figure}[h]
\begin{center}
\includegraphics[width=3.3in]{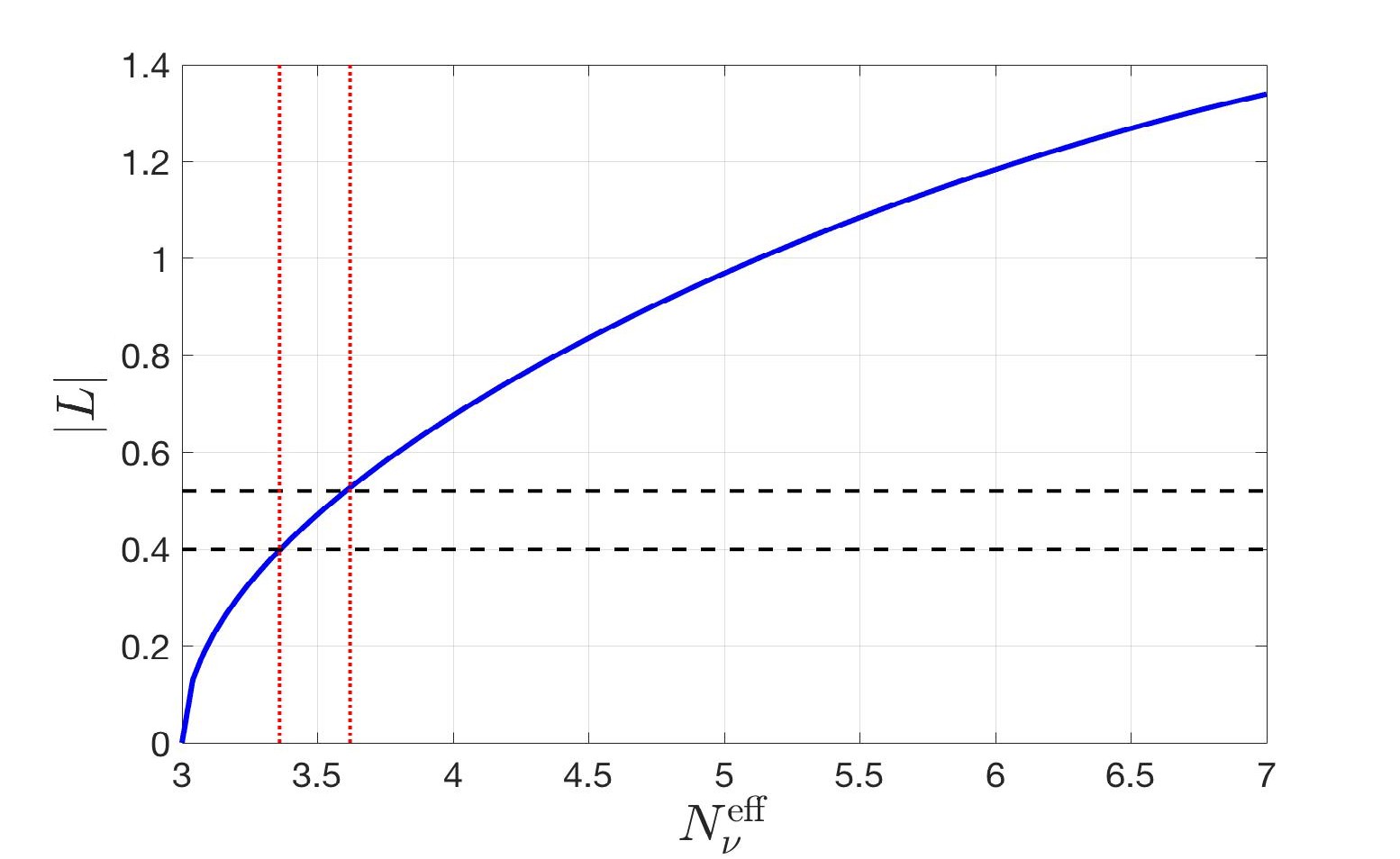}
\caption{The non-integer number of neutrino degrees of freedom $N^{\mathrm{eff}}_\nu$ resulting from lepton asymmetry in the Universe: The solid blue line shows $|L|$, the lepton-number per photon. The vertical (red) dotted lines represent the values $3.36\leqslant N_\nu^{\mathrm{eff}}\leqslant3.62$, which correspond to $0.4\leqslant|L| \leqslant0.52$ (horizontal dashed lines). }
\label{LN_Ratio}
\end{center}
\end{figure}
%%%%%%%%%%%%%%%%%%%%%%%%%%%%%%%%%%%%%%%%%%%%%%%%%%%%%%%%%%%%%%%%%%
%%%%%%%%%%%%%%%%%%%%%%%%%%%%%%%%%%%%%%%%%%%%%%%%%%%%%%%%%%%%%%%%%%
\begin{figure}[h]
\begin{center}
\includegraphics[width=3.3in]{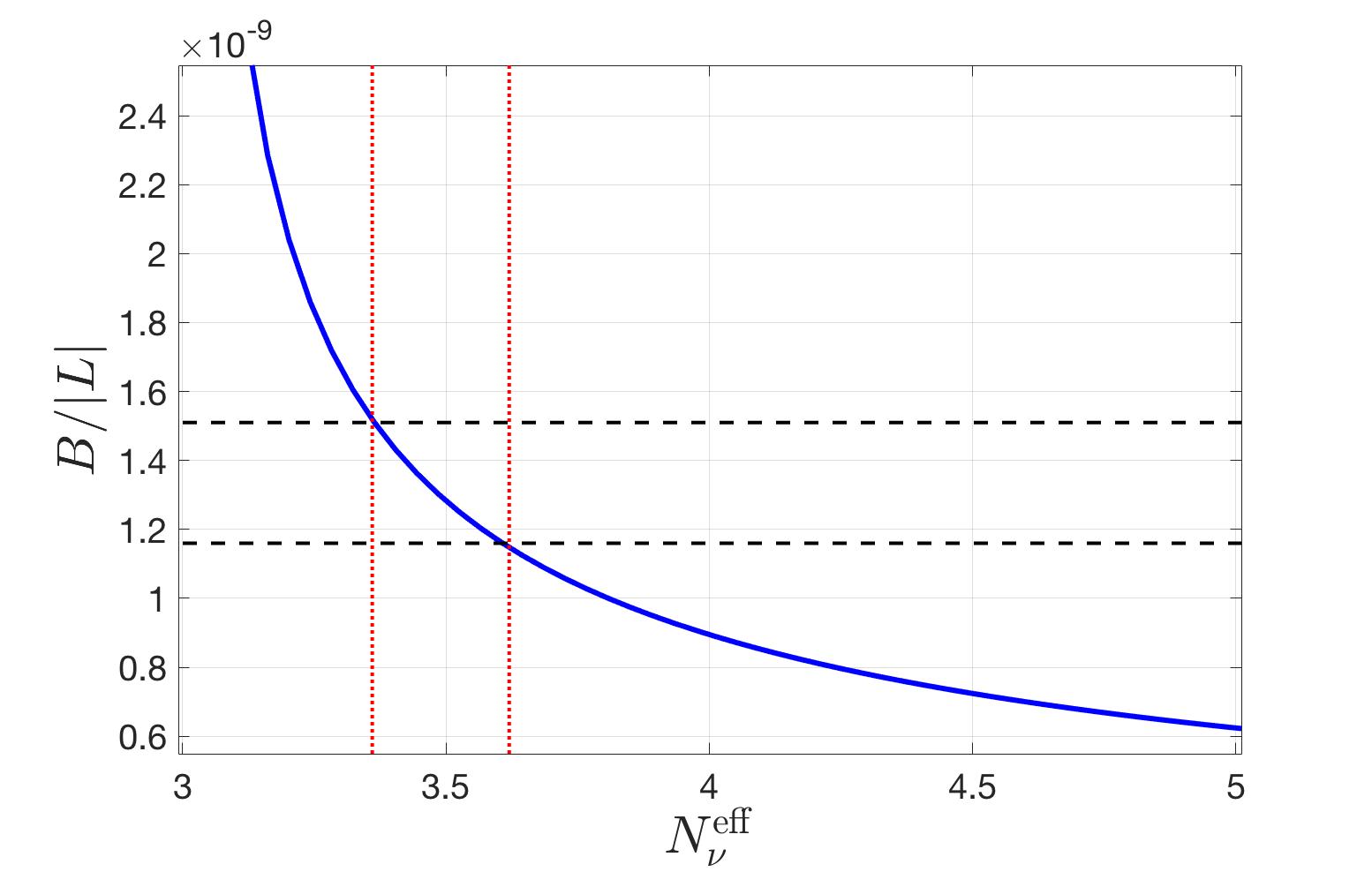}
\caption{The ratio $B/|L|$ between the net baryon number and the net lepton number as a function of $N^{\mathrm{eff}}_\nu$: The solid blue line shows $B/|L|$. The vertical (red) dotted lines represent the values $3.36\leqslant N_\nu^{\mathrm{eff}}\leqslant3.62$, which correspond to $1.16 \times 10^{-9}\leqslant B/|L|\leqslant 1.51 \times 10^{-9}$ (horizontal dashed lines).}
\label{BL_Ratio}
\end{center}
\end{figure}
%%%%%%%%%%%%%%%%%%%%%%%%%%%%%%%%%%%%%%%%%%%%%%%%%%%%%%%%%%%%%%%%%%

Finally, since the entropy-per-baryon from neutrino freezeout up to the present epoch is constant, we can obtain this value by considering the Universe entropy content today~\cite{Fromerth:2012fe}. For $T\ll1\,\mathrm{MeV}$, the entropy content today is carried in photons and neutrinos 
\begin{align}
\label{Nb_S}
\left(\frac{s}{n_B}\right)_{t_0}=\frac{\sum_i\,s_i}{n_B}=\frac{n_\gamma}{n_B}\,\bigg(\frac{s_\gamma}{n_\gamma}+\frac{s_\nu}{n_\gamma}+\frac{s_{\bar{\nu}}}{n_\gamma}\bigg)\;.
\end{align}
The entropy per particle for a massless boson at zero chemical potential is $(s/n)_{\mathrm{boson}}\approx 3.602$ and $B =n_B/n_\gamma= 0.605\times10^{-9}$(CMB) \cite{PDG:2016}. Given the nonzero neutrino chemical potential, the entropy density of the free-streaming distribution Eq.\;(\ref{fnudef}) (using the approximation $m_\nu=0$) is
\begin{align}
\label{S_N}
s_{\nu(\bar{\nu})}=\frac{4}{3}\frac{\rho_{\nu(\bar{\nu})}}{T_\nu}\mp\frac{\mu_\nu}{T_f}n_{\nu(\bar{\nu})}\;,
\end{align}
where the $-,+$ are for neutrinos and antineutrinos respectively. Substituting Eq.\;(\ref{S_N}) into Eq.\;(\ref{Nb_S}), the entropy-per-baryon can be written as 
\begin{align}
\label{Nb_S2}
\left(\frac{s}{n_B}\right)_{\!\!t_0}\!\!=\left(\frac{1}{B}\right)_{\!\!t_0}\!\!\left[\frac{s_\gamma}{n_\gamma}+\frac{4}{3T_\nu}\frac{\rho_\nu^{\mathrm{tot}}}{n_\gamma}-\frac{\mu_\nu}{T_f}\left(\frac{n_\nu-n_{\bar{\nu}}}{n_\gamma}\right)\right]_{t_0}\;,
\end{align}
where $t_0$ denotes that we take the present day values. Inserting the numerical values and using the Eq.\;(\ref{Energy_Density}) and Eq.\;(\ref{Excess_Neutrino}) we obtain the entropy-per-baryon as a function of $\mu_\nu/T_f$.

Substituting Eq.\;(\ref{delta_n}) and Eq.\;(\ref{Entropy_density}) into Eq.\;(\ref{L_B_ratio}) yields the lepton-to-baryon ratio
\begin{align}\label{L_B_ratio_final}
\frac{L}{B}=\frac{45}{4\pi^4}\frac{\pi^2\tilde\mu_\nu+\tilde\mu_\nu^3}{10.75+\frac{45}{4\pi^2}\tilde\mu_\nu^2}\left(\frac{s}{n_B}\right)_{\!\!t_0}\;,
\end{align}
in terms of $\tilde\mu_\nu=\mu_\nu/T_f$ and the present day entropy-per-baryon. 

Multiplying Eq.\;(\ref{L_B_ratio_final}) by $ B|_{t_0}$ (which then cancels in Eq.\;(\ref{Nb_S2})) we obtain the net lepton-to-photon ratio. This is shown in Fig.\;\ref{LN_Ratio}, as a function of $N_\nu^{\mathrm{eff}}$. We see that the values $N_\nu^{\mathrm{eff}}=3.36\pm0.34$ and $N_\nu^{\mathrm{eff}}= 3.62\pm0.25$ correspond to a net lepton-to-photon ratio $0.4\leqslant |L| \leqslant 0.52$ in the early universe.

In Fig.\;\ref{BL_Ratio} we show the ratio between the net baryon number and the net lepton number as a function of the effective number of neutrino species $N^{\mathrm{eff}}_\nu$ with the parameter $ B|_{t_0} =0.605\times 10^{-9}$(CMB). We find that the values $N_\nu^{\mathrm{eff}}=3.36\pm0.34$ and $N_\nu^{\mathrm{eff}}= 3.62\pm0.25$ require the ratio between baryon number and lepton number to be $1.16 \times 10^{-9} \leqslant\, B/|L| \leqslant 1.51\times 10^{-9}$. These values are close to the baryon-to-photon ratio $0.57 \times 10^{-9} \leqslant B \leqslant 0.67\times 10^{-9}$. 
%%%%%%%%%%%%%%%%%%%%%%%%%%%%%%%%%%%%%%%%%%%%%%%%%%%%%%%%%%%%%%%%%%

\section{Conclusion and Discussion}\label{Discussion}

We have addressed the tension that arises between the CMB measurement of the Planck Hubble expansion parameter $H_\mathrm{CMB}$ and current epoch measurements $H_0$. We believe that there is need for additional unobserved particles, leading to an increase in the Universe expansion rate. Considerable effort has been made in this direction, e.g., by introducing exotic and new \lq dark\rq\ particles, see~\cite{Birrell:2014cja} and references therein. 

In this work a similar effect is achieved by introducing lepton asymmetry in the Universe. This is done by introducing the chemical neutrino potential $\mu_\nu$.
Prior studies of $\mu_\nu$ explored constraints on the neutrino chemical potential by experimental data \cite{Kang:1991xa,Oldengott:2017tzj,Barenboim:2017dfq,Serpico:2005bc}: a) consequences for BBN of a large lepton asymmetry (a neutrino degeneracy) were considered, and b) Constraints on lepton asymmetry using the CMB angular power spectra were derived. In this context we believe that our work also using $\mu_\nu$, for the first time offers a quantitative connection of $\mu_\nu$ with Universe Lepton abundance and asymmetry.

The standard cosmological model assumes (arbitrarily) that the asymmetry between leptons and anti-leptons is small, similar to the baryon asymmetry; most often it is assumed $L=B$. We consider $L\simeq 1$ and explore how this large cosmological lepton yield relates to the effective number of (Dirac) neutrinos $N^{\mathrm{eff}}_\nu$. We have shown that $N^{\mathrm{eff}}_\nu>3$ results when a nonzero lepton asymmetry is hidden in the Dirac neutrino cosmic background, originating in the epoch of the neutrino freezeout from the cosmic plasma near $T=2$\;MeV. 

A neutrino abundance increase at the level of $\simeq 15$\% appears to be small. However, it reconciles tensions between the observed value of Hubble parameter in the current epoch~\cite{Riess:2018byc,Riess:2018} $H_0$ with model extrapolation $H_\mathrm{CMB}$ from the recombination epoch to the current epoch that underlie the CMB analysis. 

Fig.\;\ref{Chemical_Potential_Neff} demonstrates that it is possible to interpret the values $N_\nu^{\mathrm{eff}}=3.36\pm0.34$ (CMB only) and $N_\nu^{\mathrm{eff}}= 3.62\pm0.25$ (CMB and $H_0$) with a neutrino chemical potential $0.52\lesssim|\mu_\nu|/T_f\lesssim 0.69$. $\mu_\nu$ can have either sign, and so both $n_\nu>n_{\bar{\nu}}$ and $n_{\bar{\nu}}>n_\nu$ are possible; see Fig.\;\ref{Neutrino_Density}. In either case, such a chemical potential implies a 10.8\%-18.7\% increase, corresponding to $N^{\mathrm{eff}}_\nu=3.36$ and $N^{\mathrm{eff}}_\nu=3.62$ respectively, in the total number density of neutrinos and antineutrinos compared to the standard value, see Fig.\;\ref{Density_tot}. 

The predicted yields $|L| $ and $B/|L|$, as functions of $N_\nu^{\mathrm{eff}}$, can be seen in Fig.\;\ref{LN_Ratio} and Fig.\;\ref{BL_Ratio}, respectively. To interpret the reported values $N_\nu^{\mathrm{eff}}=3.36\pm0.34$ and $N_\nu^{\mathrm{eff}}= 3.62\pm0.25$, one needs a ratio $ 1.16 \times 10^{-9}\leqslant B/|L|\leqslant 1.51 \times 10^{-9}$. These values are within factor of two equal to the baryon-to-photon ratio~\cite{PDG:2016} $0.57\times 10^{-9}\leqslant B/N_\gamma \leqslant 0.67\times10^{-9}$. 

To conclude: motivated by the necessity to explain a slightly faster Universe expansion, we have explored the other natural scenario regarding the baryon number-to-lepton number ratio. Instead of $B\simeq |L|$, we found that $0.4\leqslant|L| \leqslant0.52$ and $B\simeq 1.33\times 10^{-9}|L|$ reconciles the CMB and current epoch results for the Hubble expansion parameter. 

%%%%%%%%%%%%%%%%%%%%%%%%%%%%%%%%%%%%%%%%%%%%%%%%%%%%%%%%%%%%%%%%%
%%%%%%%%%%%%%%%%%%%%%%%%%%%%%%%%%%%%%%%%%%%%%%%%%%%%%%%%%%%%%%%%%%

%%%%%%%%%%%%%%%%%%%%%%%%%%%%%%%%%%%%%%%%%%%%%%%%%%%%%%%%%%%%%%%%%%%
\end{document}